\documentclass[prl,twocolumn,showpacs,amsmath,letter]{revtex4}
\usepackage{graphicx}

\newcommand{\Journal}[4]{#1 \textbf{#2}, #3 (#4)}

\begin{document}

\title{Dynamical Coupling Between Ferromagnets Due to Spin Transfer Torque}
\author{Sergei Urazhdin}
\affiliation{Department of Physics, West Virginia University,
Morgantown, WV 26506}


\begin{abstract}

We use a combination of analytic calculations and numerical simulations to demonstrate that electrical current flowing through a magnetic bilayer induces dynamical coupling between the layers. The coupling originates from the dependence of the spin transfer torque exerted on the layers on the relative orientations of their magnetic moments. We demonstrate that such coupling modifies the behaviors of both layers, significantly affecting the the stability of the current-induced dynamical regimes and the efficiency of current-induced magnetic reversal.

\end{abstract}

\maketitle

Current-induced spin transfer (ST) effect~\cite{cornellorig} is the most promising mechanism for manipulation of magnetic nanodevices, due to the simplicity of the implementation and potential power benefits. The main obstacle for viable applications of the effect is the large magnitude of the required current, which is too close to the limit of the physical stability of devices. A basic magnetoelectronic device consists of a magnetic bilayer F$_1$/N/F$_2$, where F$_1$ is a magnet needed to polarize the current, N is a metallic or insulating nonmagnetic spacer, and F$_2$ is a nanomagnet whose magnetic configuration can be changed by current via ST. The efficiency of spin transfer can be characterized by the zero-temperature threshold current $I_c$ for the onset of current-induced magnetic dynamics of F$_2$. In the framework of the widely accepted spin transfer torque (STT) model~\cite{slonczewski,fert},
\begin{equation}\label{it}
I_c=eH_{eff}m_2\alpha/(\hbar g_2),
\end{equation}
where $e$ is the electron charge, $H_{eff}$ is the effective magnetic field which includes the magnetic anisotropy of F$_2$, $\alpha$ is the Gilbert damping parameter, and $g_2$ is a function of the relative orientations of the magnetic moments $m_1$ and $m_2$ of F$_1$ and F$_2$, respectively, which depends predominantly on the spin polarizing properties of F$_1$.

Several directions are pursued for reducing $I_c$. The function $g_2$ is proportional to the polarization $p=(I_\uparrow-I_\downarrow)/I$ of the current generated by F$_1$~\cite{iswvsmr}. Here, $I_\uparrow$ and $I_\downarrow$ are the spin-up and spin-down contributions to the current. Therefore, $I_c$ can be reduced by enhancing the spin-polarizing properties of F$_1$.  However, the difference between the typical value $p\approx 0.7$ for the common F$_1$ such as Py=Ni$_{80}$Fe$_{20}$ and the largest possible $p=1$ is small, limiting the room for improvement. Alternatively, $I_c$ can be reduced by decreasing $H_{eff}$, which at small external field $H$ is dominated by the anisotropy of F$_2$. Simply reducing the magnetization $M_2$ of F$_2$ would decrease $H_{eff}$, but this would also compromise the stability of the magnetic configuration. Devices with perpendicular magnetic anisotropy can in principle overcome this limitation, but the potential advantages were offset by the reduced polarization and increased $\alpha$~\cite{fullerton}. Attempts to reduce $m_2$ while stabilizing F$_2$ with an antiferromagnet have encountered similar complications~\cite{ebst}.

Here, we discuss the previously unexplored mechanism affecting the efficiency of ST, involving simultaneous current-induced effects on both ferromagnets in a bilayer consisting of F$_1$ and F$_2$. This mechanism couples the dynamics of the two magnets. We describe analytic results for a simple model system, and present realistic numerical calculations in macrospin approximation. Our results demonstrate that $I_c$ can be either increased or reduced with respect to the value given in Eq.~\ref{it} by a suitable choice of F$_1$ and F$_2$. Most importantly, we show that the dynamics of both magnets are {\it always excited simultaneously}, and thus both magnetic layers always participate in the current-induced behaviors.

\begin{figure}
\includegraphics[scale=0.6]{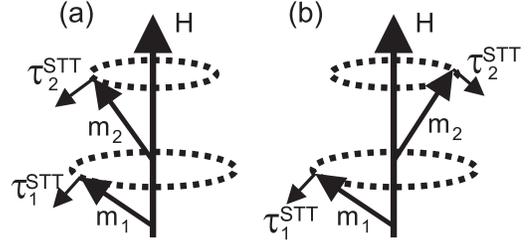}
\caption{\label{fig1} Dynamics of two magnetic moments $m_1$ and $m_2$ due to the spin torques $\tau^{STT}_1$ and $\tau^{STT}_2$ exerted on the respective moments. (a) the signs of $g_1$ and $g_2$ are the same, (b) the signs of $g_1$ and $g_2$ are opposite.}
\end{figure}

{\bf Analytic Model.} To introduce the idea of dynamical coupling of magnetic layers by STT and understand the consequences of such coupling, we consider an idealized system of two nanoscale magnets F$_1$ and F$_2$ represented by the magnetic moments $m_1$ and $m_2$, whose magnitudes are fixed in the macrospin approximation. For simplicity, this model neglects the demagnetizing fields of the ferromagnets and their dipolar coupling. The dynamics of the magnetic moments in the presence of current $I>0$ flowing from F$_1$ to F$_2$ can be described by two Landau-Lifshitz equations coupled by the STT term~\cite{slonczewski}
\begin{equation}\label{stcoupled}
\frac{dm_{1,2}}{dt}=\gamma m_{1,2}\times [H-\alpha\frac{m_{1,2}}{|m_{1,2}|}
+\frac{Ig_{1,2}}{eS_{1,2}}\frac{m_1}{|m_1|}\times\frac{m_2}{|m_2|}],
\end{equation}
where $\gamma$  is the gyromagnetic ratio,  and $S_{1,2}=m_{1,2}/(\gamma\hbar)$ are the spins of F$_1$,F$_2$. For simplicity, we assume the same $\alpha$ for both magnets, and neglect the dependence of $g_{1,2}$ on the relative orientations of $m_1$ and $m_2$.  The experimentally determined values of $I_c$ are similar in both parallel (P) and antiparallel (AP) configurations of the magnetic moments, justifying the latter approximation~\cite{iswvsmr}.  In the limit $S_1\gg S_2$, the last term in Eq.~\ref{stcoupled} for $m_1$ is negligible due to the large value of  $S_1$, resulting in a static solution for $m_1$.  The equation for $m_2$ then yields precession of $m_2$ at $I=I_{c0}$ determined by Eq.~\ref{it}, consistent with the models for the dynamics of a single ferromagnet. At $I>I_{c0}$, the AP
state becomes stable, and the solution for $m_2$ is a precessional reversal into this state.

We are interested in the solution of Eq.~\ref{stcoupled} for comparable values of $S_1$ and $S_2$. As illustrated in Fig.~\ref{fig1}, STT exerted on both layers can induce their simultaneous precession around the magnetic field $H$ with cone angles $\theta_1$ and  $\theta_2$. A simple estimate shows that a stable configuration involves precession of $m_1$ and $m_2$ with the same polar angle (Fig.~\ref{fig1}(a)). To demonstrate the stability of this configuration, we assume that $m_1$ lags behind $m_2$ by a small polar angle $\phi$. The torque ${\tau^{STT}}_1$ exerted on $m_1$ then acquires an additional component in the polar direction, resulting in the increase of its angular frequency by $\Delta\omega_1=\gamma Ig_1sin(\theta_2)sin(\phi)/[eS_1sin(\theta_1)]$. Similarly, a component of STT exerted on $m_2$ in the polar direction increases its angular frequency by $\Delta\omega_2=\gamma Ig_2sin(\theta_1)sin(\phi)/[eS_2sin(\theta_2)]$. The stability of the coupled precession with respect to fluctuations of $\phi$ requires that $\Delta\omega_1>\Delta\omega_2$, which is satisfied if $[sin(\theta_1)/sin(\theta_2)]^2<S_2/S_1$.  The derivation given below shows that the latter inequality usually holds, ensuring the stability of in-phase precession.

The stationary form of Eq.~\ref{stcoupled} for in-phase precession is
\begin{eqnarray}\label{stationary}
\nonumber Ig_1sin(\theta_2-\theta_1)=eS_1\alpha H sin(\theta_1)\\
Ig_2sin(\theta_2-\theta_1)=eS_2\alpha H sin(\theta_2).
\end{eqnarray}
The filtering properties of F$_1$ and F$_2$ are often similar, resulting in $g_1\approx g_2$. Eqs.~\ref{stationary} then lead to
\begin{equation}\label{s1s2}
S_1 sin(\theta_1)=S_2 sin(\theta_2),
\end{equation}
which satisfies the condition for the stability of in-phase precession, as discussed above. Eqs.~\ref{stationary} yield the current for the precession onset $I_c=I_{c0}/[1-S_2/S_1]$. We emphasize three main consequences of Eqs.~\ref{stationary}. Firstly, both moments {\it always precess simultaneously}.  Secondly, the relation between the amplitudes of the dynamics of the two moments is determined by the ratio $A=S_2/S_1$. In particular, the dynamics of $m_1$ is negligible at $A\ll 1$. Precession of $m_2$ can then occur only at $I=I_c>0$. Conversely, the dynamics of $m_2$ is negligible at $A\gg 1$, while precession of $m_1$ can occur at $I<0$.  Finally, $I_c$ also depends on $A$. In particular, it diverges when the magnetic moments become equal (A=1).

\begin{figure}
\includegraphics[scale=0.8]{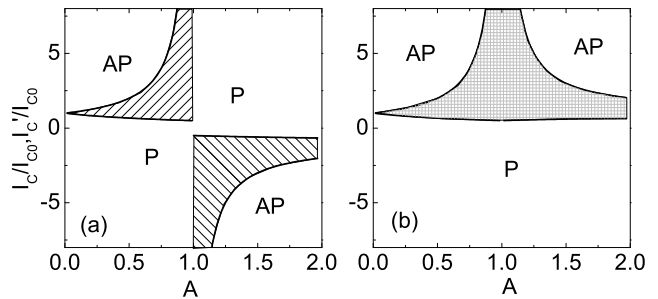}
\caption{\label{fig2} Stability diagram for the analytic model of two-layer system with varied $A=S_1/S_2$. (a) $g_1=g_2$, bistable regions are hatched. (b) $g_1=-g_2$, the region of stable precession is filled. For $A<1$, $I_{c0}$ is defined by Eq.~\ref{it}. For $A>1$, $I_{c0}$ is defined by Eq.~\ref{it} with $m_2$ replaced by $m_1$ to emphasize the dominance of this layer's dynamics.}
\end{figure}

 To determine the nature of excitations and stability regimes for the magnets coupled by STT, we insert Eq.~\ref{s1s2} into the second of Eqs.~\ref{stationary}, leading to
\begin{equation}\label{ivstheta}
I=\frac{I_{c0}/A}{\sqrt{1/A^2-1+cos^2(\theta_2)}-cos(\theta_2)},
\end{equation}
where $I_{c0}=\alpha HeS_2/g_2$. Eq.~\ref{ivstheta} describes a monotonically {\it decreasing} function of $\theta$, which implies that precession is unstable at any $I$. On the other hand, both $\theta_2=0$ and $\theta_2=\pi$ are stable in the range $I_c'=\frac{P}{1+A}<I<\frac{P}{1-A}=I_c$, where $I_c'$ is the current at which the AP state becomes stable.  The resulting stability diagram is shown in Fig.~\ref{fig2}(a), where the bistable regions are hatched.

As a consequence of the dynamical coupling due to STT, the value of $I_c'$ is reduced by a factor of two at $A=1$, while $I_c$ diverges. It remains to be seen whether these coupling effects can be utilized to increase the efficiency of magnetoelectronic devices. Current-induced bistability in devices with $A\approx 1$ should generally result in telegraph-type noise due to thermally activated transitions between the two stable configurations. Such noise is detrimental to most applications. However, such configuration may be useful if small-amplitude current-induced dynamics is undesired.  For $A<1$, dynamical coupling also makes it impossible to induce dynamical states at $I<0$. Similarly, dynamics cannot be induced at $I>0$ for $A>1$. We shall see below that this result also holds for a more realistic model. The fact that bipolar current-induced excitations were observed~\cite{cornellbipolar,kentbipolar} must indicate a significant  breakdown of the macrospin approximation in these experiments.

 Applications of magnetoelectronic devices in microwave generation require $I_c$ to be minimized, and stable precession to be achieved over a significant range of $I$.  We demonstrate below that $I_c$ can be reduced by the current-induced coupling when $g_1$ and $g_2$ have opposite signs. This requires reversing the spin anisotropy of one of the ferromagnets, which can be accomplished by doping F$_1$ or F$_2$ with impurities providing appropriate spin-dependent scattering in their bulk, and/or by choosing N that inverts the spin anisotropy of electron scattering at F/N interface~\cite{inverted}. For simplicity, we now assume that $g_1=-g_2<0$. The torques exerted on the two magnets now result in the mutual attraction of the moments at $I<0$, resulting in a stable collinear configuration at $I<0$ for any $A$. Dynamical states are induced only by $I>0$, regardless of the value of $A$. Fig.~\ref{fig1}(b) shows that the relative precession phases of $m_1$ and $m_2$ are now shifted by $180^0$. Estimates similar to the ones presented above for $g_1,g_2>0$ show that this precessional configuration is stable.

  Eq.~\ref{s1s2} for the precession cone angles of the two moments still holds, but Eq.~\ref{ivstheta} is replaced with
  \begin{equation}\label{ivstheta2}
I=\frac{I_{c0}/A}{\sqrt{1/A^2-1+cos^2(\theta_2)}+cos(\theta_2)}.
\end{equation}
The resulting stability diagram shown in Fig.~\ref{fig2}(b) includes a region of stable coupled precession of the two magnets.  At $A=1$, the precession onset current is reduced by a factor of two, while the largest value of $I$ at which the precession remains stable diverges.

{\bf Numerical Simulations.} To determine whether the current-induced coupling remains robust in realistic systems, we solved Eqs.~\ref{stcoupled} numerically. We used the geometry and the magnetic properties typical for the spin-transfer devices, in which F$_1$ and F$_2$ are Py layers with $M=800$~emu/cm$^3$ and thicknesses $t_1$ and $t_2$, patterned into an elliptical shape with dimensions $120\times 60$~nm~\cite{cornellorig}. We used the material parameters known from the magnetotransport measurements~\cite{bassgmr}.  The demagnetizing fields were taken into account. The dipolar coupling was neglected to eliminate its influence on the interpretation of current-induced behaviors.  We included the dependencies of the spin transfer efficiencies $g_1,g_2$ on the configuration of the magnetic system to take into account the slight experimental asymmetry between the P$\rightarrow$AP and AP$\rightarrow$P reversals~\cite{iswvsmr}.  The equations were numerically integrated by the stochastic Heun method with a fixed time step set to $2$~psec. To verify the convergence, the step size was decreased by a factor of two, which  did not significantly affect the results.  Random field approximation was used to model the thermal activation between different current-induced modes, with temperature set to $50$~K. Damping parameter $\alpha=0.03$ was used for both magnets~\cite{cornellscience}.

\begin{figure}
\includegraphics[scale=0.9]{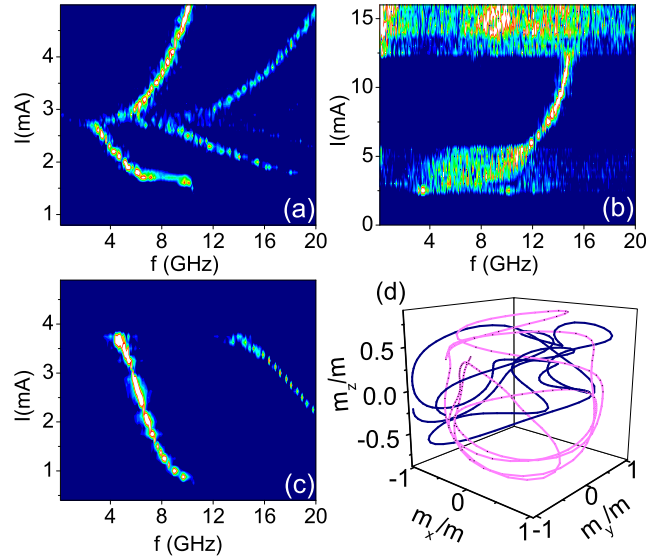}
\caption{\label{fig3} (a)-(c) Normalized spectral intensity for the dynamics of the in-plane hard-axis component of $m_2$, (a) $A=0.2$, (b) $A=0.67$, (c) $A=1$ and $g_1=-g_2$. Lighter colors correspond to higher intensity. The same scale is used in all three plots. (d) Time trace for the trajectories of $m_1$ (dark blue) and $m_2$ (light pink) for $A=0.67$, $I=15$~mA. $x$ is the easy direction, $y$ is the hard in-plane direction, and $z$ is perpendicular to the film plane. The calculations were performed at $H=1$~kOe along the easy axis, and a fixed $t_2=4$~nm.}
\end{figure}

The model was tested for three limiting cases: $t_1\gg t_2$, $t_1\ll t_2$, and $t_1=t_2$. In the first case, STT induced the dynamics of only $m_2$, and only at $I>0$. Fig.~\ref{fig3}(a) shows the normalized spectra of the component of $m_2$ along the hard in-plane axis for $A=0.2$. The peaks correspond to several harmonics of the precession frequency.  After the onset of small-angle precession at $I_c=1.2$~mA, the frequency of precession decreases due to transition to large-angle clamshell precession trajectory. The transition to the out-of-plane precession at $I>2.6$~mA results in the increase of precession frequency. These results are consistent with the published calculations for single layer dynamics, and are in overall agreement with the experiments~\cite{kiselevnature}. Similar spectra were obtained for the dynamics of $m_1$ at $I<0$, for $t_1\ll t_2$. Calculations for $t_1=t_2$ yielded identical spectra for $m_1$ and $m_2$ regardless of the sign of $I$, as expected for this symmetric geometry. Additionally, the calculations for $t_1=t_2$ and $H=0$ reproduced the sequential thermally activated flipping of $m_1$ and $m_2$ experimentally observed in symmetric magnetic nanopillars~\cite{wenglee}.

Current-induced coupling of the dynamics of two layers significantly affected the spectra at $0.4<A\le 1$, as illustrated in Fig.~\ref{fig3}(b) for $A=0.67$. The onset current for the magnetic dynamics is larger than in Fig.~\ref{fig3}(a), in agreement with the results of the analytic model presented above. At $3$~mA$<I<6$~mA, the data exhibit a broad incoherent excitation spectrum due to random transitions between different types of precession dynamics of $m_2$.  Specifically, the precession of $m_2$ alternates between the clamshell-type and the out-of-plane trajectories, depending on the relative phase with the elliptical precession of $m_2$. The latter is not phase-coherent with $m_2$. The formation of a sharp peak at $6$~mA$<I<12$~mA is associated with the complete transition to out-of plane precession of $m_2$, resulting in nearly static deflection of $m_1$ towards $m_2$.  At $I>12$~mA, the oscillations of $m_1$ become sufficiently large to disrupt the periodic precession of $m_2$, resulting in increasingly chaotic dynamics of both. Fig.~\ref{fig3}(b) shows the trajectories of $m_1$ and $m_2$ over a 0.5~nsec period at $15$~mA. Despite nearly chaotic behavior of both moments, some parts of the trajectory of $m_2$ resemble clamshell and out-of-plane precession. The chaotic behaviors persisted in deterministic simulations at $T=0$, eliminating thermal fluctuations as their cause. Based on the analysis of the trajectories, we believe that these behaviors are caused by a combination of large phase space associated with the dynamics of
both magnets, and nonlinear nature of the large-amplitude magnetic dynamics.

At $A>0.67$, the broad excitation band at small $I$ and the out-of-plane precession peak gradually disappeared. At $A=1$, only the $I>12$~mA continuum remained, and similar features appeared at $I<-12$~mA. There were no excitations at $I<0$ for all $A<1$, in agreement with the analytic results described above. These results seem to be inconsistent with the intuitive picture, according to which $I>0$ can induce the dynamics of $m_2$, while $I<0$ can induce the dynamics of $m_1$ even for $A<1$~\cite{tsoisymmetric}. However, analysis of the magnetic trajectories shows that all oscillations of $m_1$ at $I<0$ are efficiently suppressed by $m_2$ closely following $m_1$, thus reducing STT exerted on both layers.

Calculations for $g_1=-g_2$ showed that $I_c$ is reduced in this case, in agreement with the analytic model. Fig.~\ref{fig3}(c) shows spectra calculated for $A=1$. $I_c=0.75$~mA is slightly over half of the value $I_{c0}$ obtained for $A\ll 1$. Therefore, reduction of the critical current for $g_1=-g_2$ appears to be a robust feature of dynamical coupling. The spectra do not exhibit broadening and chaotic dynamics characteristic of the $g_1=g_2$ data. The peaks in Fig.~\ref{fig3}(c) abruptly terminate at $I=3.8$~mA due to the formation of a static stable configuration with $m_1$ and $m_2$ oriented opposite to each other perpendicular to the film plane. Such static configuration is, however, not present for $A<1$.

We have performed additional calculations without some of the simplifying assumptions of our model. In particular, we tested the effects of different magnetic damping in the two magnets and different magnetizations. We also checked the effect of the dipolar coupling between the layers. The modifications did not qualitatively change our conclusions regarding the effects of current-induced coupling, as long as the resonant frequencies of the two layers remained similar to each other.

In summary, we have analyzed the simultaneous effects of spin torque on both layers in a magnetic bilayer. We showed that the spin torque results in dynamical coupling between the layers, modifying their individual current-induced dynamical properties. In particular, the onset of magnetic precession of both magnetic always occurs at the same current. In case of similar spin-transport anisotropy of both layers, the onset current for magnetic precession is increased, diverging for identical values of the moments of the two layers. Realistic numerical simulations show that coupling often leads to incoherent dynamical regimes associated with transitions between different dynamical states of the bilayer. However, for the opposite spin-transport anisotropy of the two layers, the onset current for the magnetic precession is decreased, and coherence is maintained even for the identical dimensions of the magnetic layers. This effect of dynamical coupling on the spin transfer efficiency can become useful for implementing devices with improved performance.

This work was supported by the NSF Grant DMR-0747609. I thank Lidia Novozhilova for help with numerical simulations.

\end{document}